\documentclass{article}
\usepackage[english]{babel}
\usepackage[latin1]{inputenc}
\usepackage{graphicx}
\usepackage{amssymb,amsmath}
\usepackage{geometry}
\usepackage{pstricks}
\usepackage{dcolumn}
\usepackage{bm}
\title{Formation of spanwise vorticy in oblique turbulent bands of transitional plane Couette flow, part 1: Numerical experiments.}
\date{{\small Received: 6 February 2014, Accepted in EJMBF: 10 November 2014\\
DOI: 10.1016/j.euromechflu.2014.11.007}}

\author{Joran Rolland\footnote{joran.rolland@ladhyx.polytechnique.fr} \footnote{LadHyX, UMR CNRS 7646, \'Ecole Polytechnique, 91128 Palaiseau, France ;
 INLN, UMR CNRS 7335, UNSA, 1361 route des lucioles, 06560 Valbonne, France.}}

\begin{document}

\maketitle

\begin{abstract}
  This article investigates the formation of spanwise vorticity in the velocity streaks of the oblique laminar-turbulent bands of plane Couette flow (PCF) by mean of Direct Numerical Simulations (DNS). The spanwise vorticity is created by a roll--up type development of the streamwise-wall normal shear layer of the velocity streaks. It is advected by the large scale flow along the bands. We propose a criterion on spanwise vorticity which detects these events in order to perform systematic measurements. Beside of the streamwise and spanwise correlation lengths of the rolls, their advection velocity is measured and shown to match  the large scale flow along the band near the turbulent region. Eventually, we discuss the possible relation between ejection of vorticity away from the bands near the laminar region and the size of said laminar region.
\end{abstract}

\begin{flushleft}{\small \emph{Keywords:} transition, Wall-bounded turbulence, shear flows instabilities

}\end{flushleft}

\section{Introduction}

\begin{figure}
\centerline{\includegraphics[height=3cm]{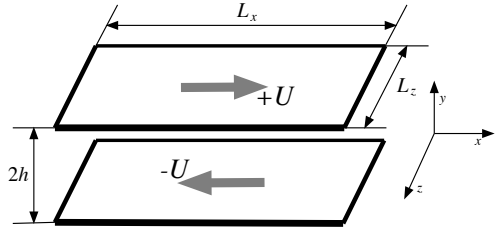}}\caption{Sketch of plane Couette flow indicating the parameters and coordinate system.}\label{f1}\end{figure}

\begin{figure}
\centerline{{\large
\textbf{(a)}\hspace{1mm}\includegraphics[width=4cm,clip]{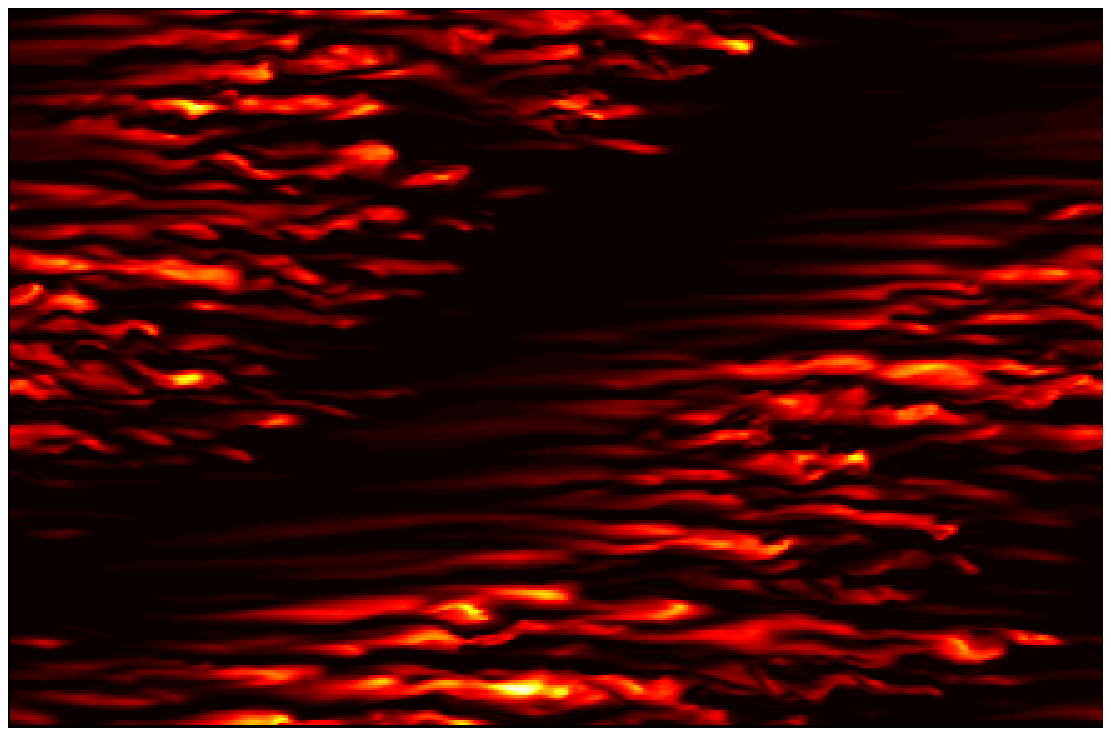}\hspace{1mm}\textbf{(b)}\hspace{1mm}}\includegraphics[width=4cm,clip]{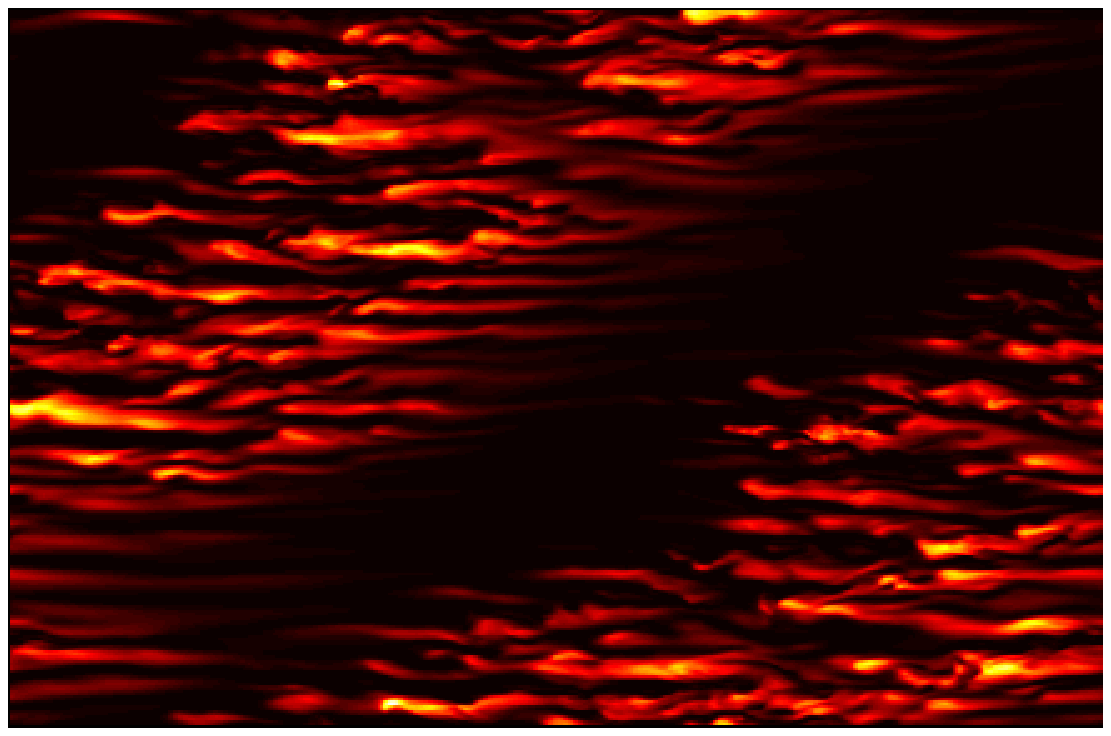}}
\caption{Example of turbulent oblique band, colour plot of
$\bf{v}^2$ (a): in a $y=-0.62$ plane, (b) : and $y=0.62$ plane. Computed by DNS in a
periodic domain, $L_x=110$, $L_z=72$, $R=370$. }\label{f1_}
\end{figure}

    This paper is focused on the statistical
analysis of a secondary instability of turbulent streaks in transitional plane Couette flow. Plane Couette flow (PCF), the flow between two parallel moving
planes (sketched in figure~\ref{f1}), displays a discontinuous transition to
turbulence. The laminar baseflow is linearly stable for all Reynolds
numbers $R$ and can coexist in space and in time with turbulent
flow. Turbulence can be sustained above a first Reynolds number
$R_{\rm g}\simeq 325$, and takes the form of oblique turbulent bands
\cite{prigent,RM,BT07,BT11} (Fig.~\ref{f1_}) which correspond to a modulation of
turbulence \cite{RM,BT07}.  All laminar troughs disappear
above a second Reynolds number $R_{\rm t}\simeq 415$.
Other wall-bounded flows like plane
Poiseuille flow display oblique laminar-turbulent coexistence \cite{ATK}.
Poiseuille pipe flow (PPF) also displays coexistence of laminar and turbulent
flow as well in an unsteady manner : the low Reynolds number puff regime displays relaminarisations and splitting, while turbulence invades the whole pipe in the high Reynolds number slug regime \cite{avila,DWK,SK}.

  From a microscopic point of view, the turbulent regime beyond $R_{\rm t}$
 is well understood in term of the self sustaining
process of turbulent streaks and streamwise vortices \cite{W,schhu}.
The turbulence inside the bands consists of velocity streaks and streamwise
vortices as well \cite{BT07} (Fig.~\ref{f1_}). However, the self-sustaining process of wall-bounded
turbulence alone is insufficient to explain the coexistence of laminar and turbulent flow at low Reynolds number. Several results point toward possible mechanisms. Using DNS of the flow, one can show that the wavelengths of the bands are related to the Reynolds number through the balance of time averaged diffusion and advection in the laminar part of the flow \cite{BT07}. DNS, Coupled map and reaction-diffusion models \cite{BPPF,B,PRL} brought insight on the role of advection of small scale chaos by the large scale coherent flow in order to turn local transient chaos into global sustained turbulence.

  In this paper, we seek to determine the phenomena behind these models and averaged results. A specific
activity found in the trailing edges of puff and slugs of pipe flow \cite{DWK,SK} provides a starting point. One can see formation
of azimuthal vorticity, \emph{via} a destabilisation of the shear layer of slow speed streaks. The vorticity is advected toward (puff regime) or away from (slug regime) the turbulent zone and feeds turbulence. This led to the suggestion of a self-sustaining process of the puffs \cite{SK}.
Besides, it was argued that the advected vorticity was accountable for the expansion of slugs
\cite{DWK}, provided the speed of the vortices is lower than the
advection velocity of the slug. The study of a slightly idealised shear layer indicated that the vorticity formation mechanism is along the lines of a Kelvin--Helmholtz instability \cite{SK}. A similar type of activity can be found in the leading edge of puffs, due to the inflectional nature of the velocity profile \cite{HDAS}. Aida \emph{et al.} \cite{ATK} reported a comparable
phenomenon in plane Poiseuille flow in the spot regime.

\begin{figure}
\begin{flushleft} \textbf{(a)}\end{flushleft}
\centerline{\includegraphics[height=5cm,clip]{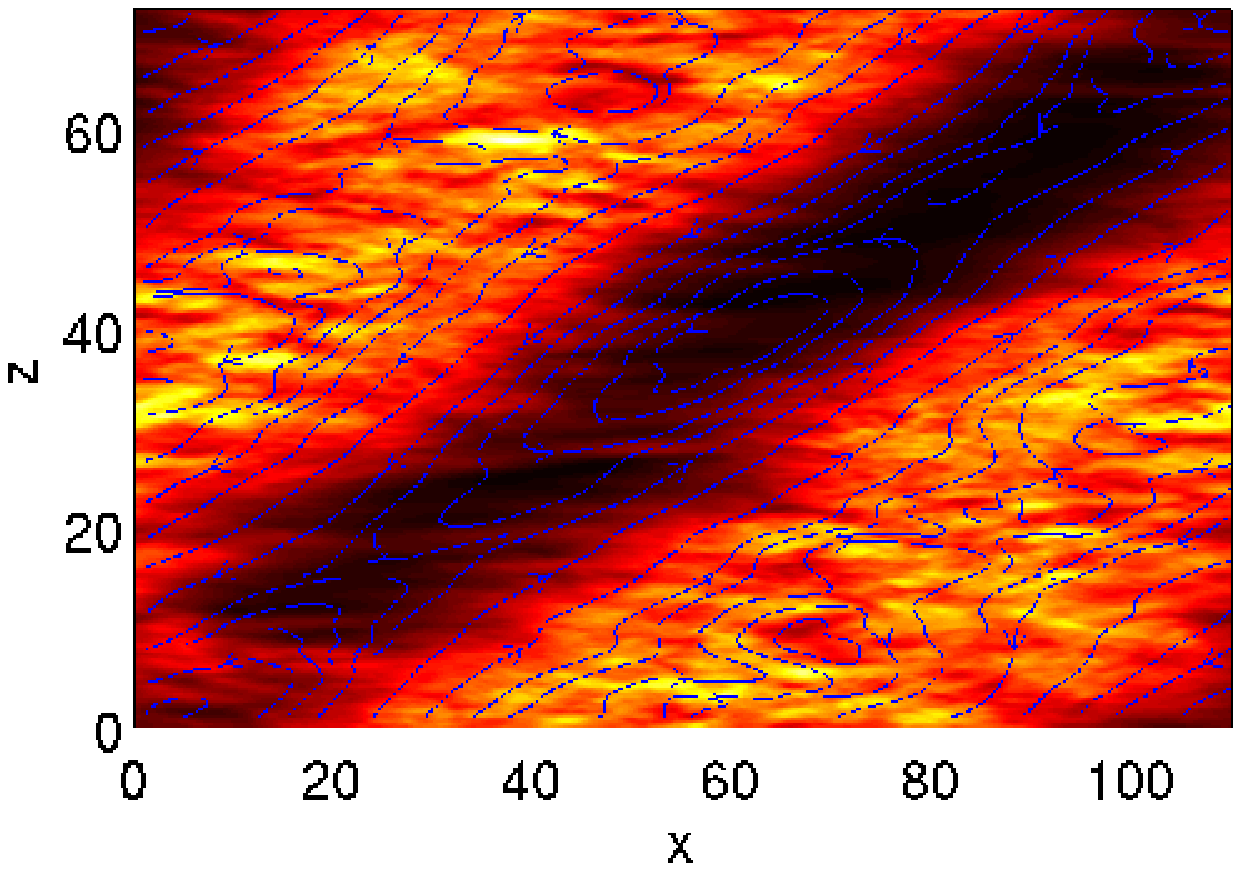}\hspace{1cm}
\begin{pspicture}(7,7)
\rput(0,6.5){{\Large \textbf{(b)}}}
\psline{}(0.5,0.5)(0.5,6.5)
\psline{}(0.5,6.5)(6.5,6.5) \psline{}(6.5,6.5)(6.5,0.5)
\psline{}(0.5,0.5)(6.5,0.5) \psline[linecolor=gray]{}(0.5,2)(5,6.5)
\psline[linecolor=gray]{}(2,0.5)(6.5,5) \rput(3.5,4){$L$} \rput(3.25,5){$I$}
\rput(3.75,2){$I$}
 \rput(2.25,5.5){$T$} \rput(5.25,1.5){$T$}
\psline{->}(1.75,4.75)(1.45,4.45)
\psline{->}(2,4.5)(1.25,3.75)
\psline{->}(2.25,4.25)(0.75,2.75)
\psline{->}(2.5,4)(1.75,3.25)
\psline{->}(2.75,3.75)(2.45,3.45)
\psline{->}(4.25,2.25)(5.75,3.75)
\psline{->}(4.5,2)(5.25,2.75)
\psline{->}(4.75,1.75)(5.05,2.05)
\psline{->}(4,2.5)(4.75,3.25)
\psline{->}(3.75,2.75)(4.05,3.05)
\psline[linecolor=gray]{}(0.5,5.6)(1.4,6.5)
\psline[linecolor=gray]{}(5.6,0.5)(6.5,1.4)
\end{pspicture}
}
\caption{Large scale flow around the band. (a):  Resulting flow along the
bands. The streamlines indicate wall normal averaged streamwise and spanwise
velocity fields. The colour levels indicate wall normal averaged norm of
velocity $\mathbf{v}^2$ (DNS result). (b) : sketch of that flow, noting the Turbulent and Laminar zones as well as the Intermediate zones between the two.}
\label{2d2c}
\end{figure}

  Similar short wavelength instabilities have been pointed out in the growing spots of PCF \cite{ispspot}.
They lead to a formation of spanwise vorticity that differ from the tongues of spanwise vorticity associated
with the self sustaining process of turbulence \cite{schhu}. A preliminary study in an idealised situation indicates that, again, the vorticity formation mechanism is very likely a Kelvin--Helmholtz instability \cite{ETC}. It was shown that the quadrupolar flow around the spot
advected the spanwise vorticity toward the edges where they very likely contribute to the extension of the spot.
The advection of such vorticity may very well be central to the sustainment mechanism of the steady bands as well.
Unlike PPF or the spots of PCF, the bands have a steady large scale flow. Its full three dimensional,
three components structure has first been computed in DNS by Barkley \& Tuckerman by averaging in time over $2000$ eddy turnover time \cite{BT07}.
A final wall normal average shows a circulation along the bands \cite{BT07,M11} (Streamlines in figure.~\ref{2d2c} (a), Sketched in figure~\ref{2d2c} (b)),
which varies sinusoidally over a scale of fifty half gaps.
This large scale flow is strongest in the Intermediate (or overhanging \cite{cole66}) zone between laminar and turbulent flow. This zone
is somewhat equivalent to the trailing edge of puffs and slugs.
Advected rolls in the intermediate zone may therefore constitute a starting point for the study of the sustainment of the bands.

  In order to address this question, the article is organised as follow. The first section contains a description of the system and a reminder of our procedure (\S\ref{sys}). The roll formation in the
velocity streaks is displayed
(\S~\ref{inst}). The link to spanwise vorticity is then explained. The measurements, autocorrelation function and
advection velocity of perturbations are then considered in \S\ref{mes}.
 These results are eventually summed up and
discussed in the last section
(\S\ref{concl}).

\section{Numerical procedure\label{sys}}

  The system studied is plane Couette flow, with periodic in plane boundary conditions (Fig.~\ref{f1}).
The velocity of the moving plates placed at $y=\pm h$ is $\pm U$.
 These quantities are used to make dimensionless
velocities ($\mathbf{v}/U$), length ($\mathbf{x}/h$) and time ($th/U$). The Reynolds number $hU/\nu$, with $\nu$ the kinematic viscosity, is
the control parameter for the transition in large systems \cite{prigent,RM,BT11}. Together with the sizes $L_x$, $L_z$
(Fig.~\ref{f1}), they set the whole statistical behaviour of the
flow \cite{RM,BT11,PM}. In addition to the coordinate system $(x,y,z)$, we call $z'$ the direction along the band. The flow field can be written $\mathbf{V}=y
\mathbf{e}_x+\mathbf{v}$ where $y \mathbf{e}_x$ is the laminar
baseflow and $\mathbf{v}$ the departure from the laminar baseflow. The norm of the departure $\mathbf{v}^2$ is called the energy.

The incompressible Navier--Stokes equations are numerically integrated
using J. Gibson's Direct Numerical Simulation code {\sc
channelflow} \cite{gibs}. More details on our implementation and use
of the code can be found in previous articles \cite{RM,MR}. The in-plane resolution is $N_{x,z}/L_{x,z}=4$
(using the $2/3$ de-aliasing rule) and wall normal resolution is
$N_y=27$. This resolution ensures reliable
quantitative results \cite{dsc10}.

  The oblique bands
regime (Fig.~\ref{f1_}) is found between Reynolds numbers $R_{\rm g} =325\pm 5$ and
$R_{\rm t}=410\pm 5$. This is in agreement with previous numerical simulations and experiments \cite{prigent,BT07}. A domain containing one band is sufficient to
study the roll formation in PCF. Therefore the data presented in the band case
is at $R=370$, in the middle of the range $[R_{\rm g}; R_{\rm t}]$, in a domain of size $L_x=110$ and $L_z=72$, which is optimal to accommodate one band.

The band regime is obtained by the following procedure: a smooth random velocity field is integrated in time for a duration of $500h/U$ at $R=500$,
in order to obtain uniform wall turbulence. The Reynolds number is then decreased at $R=370$,
and the velocity field in integrated in time for a duration of $1500h/U$ in order to reach the steady band regime.
This velocity field is used as an initial condition to perform the study of the vorticity formation.
The flow is statistically steady and the lifetime of turbulence is tremendously long at this Reynolds number \cite{shi},
therefore, the same results can be obtained after any time integration of the initial condition.

\section{Spanwise vorticity formation: microscopic description in bands\label{inst}}

In this section we first verify that departure to the velocity streaks can be viewed as instabilities (\S~\ref{frame}). Then, the roll formation in the velocity streaks is identified \emph{via} two dimensional visualisation of DNS (\S~\ref{visu}). In order to go beyond visualisations, we use spanwise vorticity to characterise the rolls
(\S\ref{vort}), from which we define a marker of the rolls.

\subsection{A statistically steady and coherent flow \label{frame}}

\begin{figure}
\centerline{\includegraphics[width=6.5cm,clip]{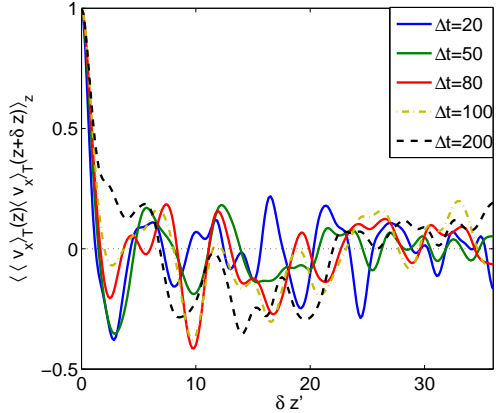}}
\caption{Autocorrelation function of the time averaged streamwise velocity $v_x$ along a diagonal $z'$, as a function of $\delta z'$, for increasing averaging times, sampled in a statistically steady band.}
\label{f4bf}
\end{figure}

We will investigate a possible secondary instability of the streamwise vorticity streaks. As it is often the case in the study of secondary instabilities in wall normal turbulence \cite{SK,W,schhu}, we will consider the streaks to be frozen, since they are evolving on a long time scale. In order to test this approximation, one can compute the normalised autocorrelation function of the streamwise velocity averaged over a duration $\Delta t$ (Fig.~\ref{f4bf}). The autocorrelation function is computed as a function of $z'$, the diagonal direction of the band. This coordinate is defined such that $\mathbf{e}_{z'}=1/(\sqrt{L_x^2+L_z^2})(L_x\mathbf{e}_x+L_z\mathbf{e}_z)$. On each diagonal starting at $x=x_0,z=0$, one has $z'=\sqrt{(x-x_0)^2+z^2}$.

One can note two interesting facts in the spirit of the frozen streaks framework. Firstly, the flow is nearly periodical and coherent over a long distance: this is shown by the modulation of the correlation function and the slowly decreasing envelope. The first minimum is approximately at $\delta z'=2$ and the modulation has a non negligible amplitude over a distance $\delta z'\gtrsim 20$. Secondly, the correlation function is time invariant and the flow remains coherent over long period of time: coherence starts to disappear for averaging over durations larger than $\Delta t=100$. This result is not surprising, since the turbulent bands are merely a long wavelength modulation of the classical low Reynolds number velocity streaks. This gives a firm basis to the preliminary linear analysis \cite{ETC}.

\subsection{A Typical example \label{visu}}

\begin{figure}
\centerline{\includegraphics[width=17cm,clip]{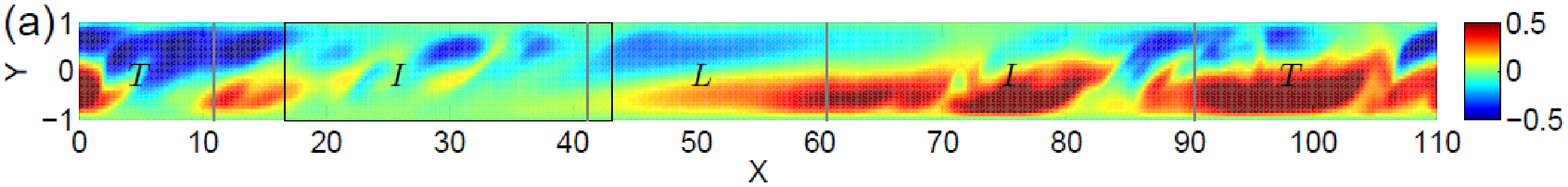}}
\centerline{{\large\textbf{(b)}\hspace{0.1cm}\includegraphics[width=3cm,clip]{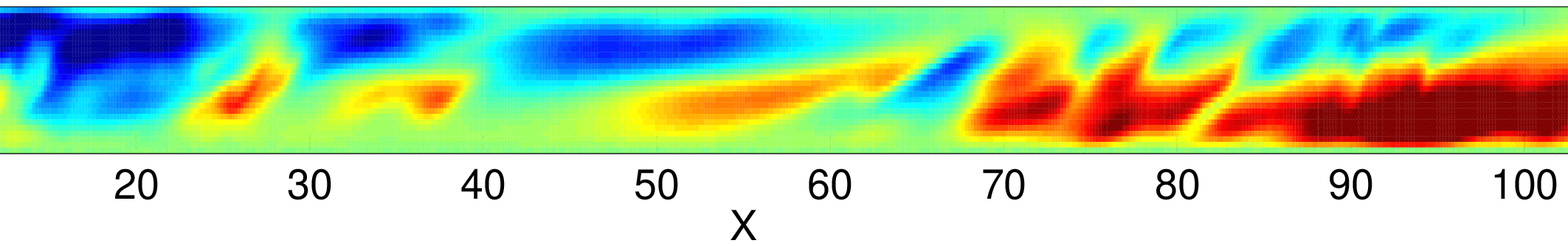}\hspace{0.1cm}\textbf{(c)}\hspace{0.1cm}
\includegraphics[width=3cm,clip]{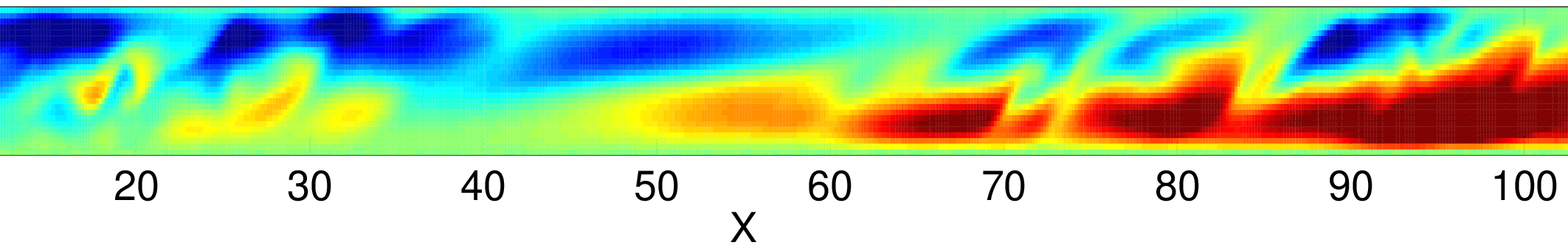}\hspace{0.1cm}\textbf{(d)}\hspace{0.1cm}
\includegraphics[width=3cm,clip]{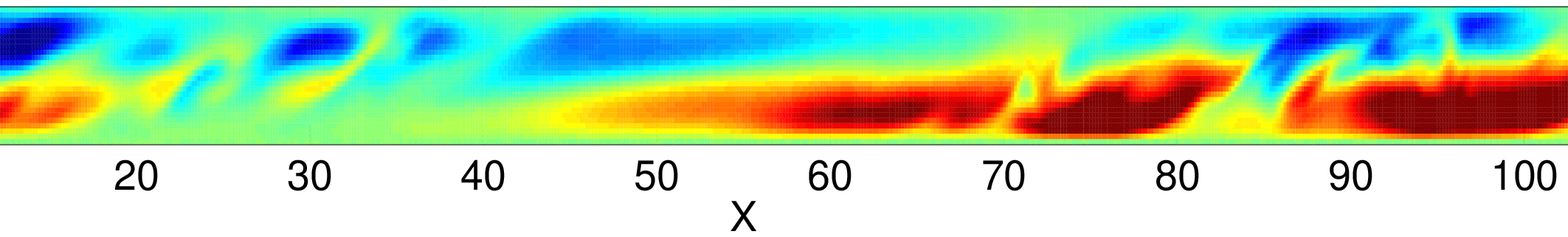}\hspace{0.1cm}\textbf{(e)}}\hspace{0.1cm}
\includegraphics[width=3cm,clip]{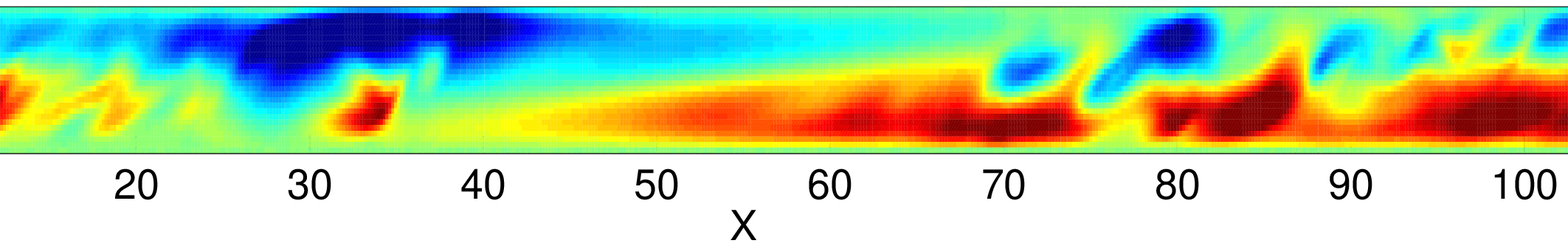}}
\caption{Colour levels of the streamwise velocity field in a $z=cte$ plane at successive instants.
(a): $T=20$. The Intermediate, Turbulent and Laminar zones are indicated by letters I,T and L and separated by gray lines. Zoomed on the perturbation (b): $T=0$, (c): $T=10$, (d): $T=20$, (e): $T=30$.}
\label{f2}
\end{figure}

  We present a typical example of the roll occurring in an intermediate zone of the band. The turbulent bands are obtained by our procedure, and are followed in time until clear realisations of the events are seen. Colour levels of the streamwise velocity field in a $x-y$ plane at
successive instants are displayed in figure~\ref{f2}. The whole streamwise length and gap are included (Fig.~\ref{f2} (a)), and one can see the turbulent zone at $90\lesssim x\lesssim 110$, $0\lesssim x \lesssim 10$, and the intermediate (or over-hanging zones \cite{cole66}) at $60\lesssim x\lesssim 90$ and $10\lesssim
x\lesssim 40$. For the most part of the flow, one can see the typical velocity streaks of plane Couette flow in the long scale ($L_x=110$) modulation of the bands.

However, an intermediate region ($15< x <45$) departs strongly from the shear layer seen in the other intermediate region (transformed by the centro-symmetry of the band \cite{BT07}) and displays the same type of roll-up as the slow speed streaks of pipe flow \cite{DWK,SK}. Unlike what is seen in pipe flow, the roll is centered around $y=0$ and extends over the whole gap. Like pipe flow, this roll in concentrated in the $z$ direction.  The two rolls have a typical size of $5h$, leading to a wavevector of order $1$.

Setting an origin of time $20h/U$ before the snapshot of figure~\ref{f2} (a), one can follow the behaviour of the rolls in the $15<x<45$ frame by snapshots every $10 h/U$. We can first see the expected shear layer ($T=0$ Fig.~\ref{f2} (b)). A portion of roll appears in the frame (Fig.~\ref{f2} (c)), then followed by the full rolls (Fig.~\ref{f2} (d)). The rolls then disappear (Fig.~\ref{f2} (e)) in a duration much shorter than the typical viscous decay time (of order $1/R$). There is no apparent effect of vorticity stretching, this change is most likely an effect of the advection of the roll. The advection appears much more clearly in the $z$ direction than it does in the $x$ direction.

Note that in unconstrained DNS of turbulent flows such as ours, the study is limited to the non-linear development of the roll \cite{DWK,SK}. This limitation has manifested itself in the former studies of coherent structures of shear flows turbulence \cite{Jimenez91}. Only constrained simulations \cite{HKW} or idealised models \cite{W} can shed light on the onset of the rolls formations. In the case of azimuthal vorticity formation in pipe flow, the examination of an idealised version of the shear layers found in DNS confirmed that a Kelvin--Helmholtz instability was at the source of the vorticity formation. In the case of plane Couette flow, a preliminary study of an idealised shear layer indicated that a kelvin--Helmholtz instability is also the likely cause of the roll formation \cite{ETC}. A full fledged stability analysis will be proposed in the second part of this article \cite{isp2}. Such a mechanism should be expected: other instabilities are very unlikely.

\subsection{Spanwise vorticity as a marker \label{vort}}

  In order to go beyond the visualisations and perform measurements on these rolls (lengthscale, advection velocity \emph{etc}.), we quantitatively justify the use of a marker derived from the spanwise vorticity $\omega_z=\partial_x (v_y)-\partial_y (v_x)$. This marker has already be used in the study of such rolls in the spots \cite{ispspot}.

\subsubsection{Principle}

  We start from our example. Colour levels of
$\omega_z$ in the same $x-y$ plane as figure~\ref{f2} (a) are
displayed in figure~\ref{f3} (a), for comparison with the
velocity field. The framed region is the same as in figure~\ref{f2}
(a,d). A general view of the flow shows that nearly everywhere, there
is $\omega_z<0$ near the walls and $\omega_z>0$ in the core region. This is the spanwise vorticity field expected for the velocity streaks. We can see a few tongues of $\omega_z<0$ going from the walls to the center of the gap, related to the self sustained process of turbulence \cite{schhu}. The framed region, however, displays $w_z<0$ in the mid gap, where $w_z>0$ is expected. We will take advantage of this fact to build our marker.

\subsubsection{The coherent vorticity field}

So as to generalise this observation, we first measure the average vorticity profiles in each region (laminar, turbulent, intermediate), that will give us the vorticity profiles in the velocity streaks. This will give a quantitative description of the background on which the negative vorticity stands out.

For that matter, we use a discrimination method between laminar and turbulent flow that allows us to build a spatial mask $I^{\rm t,l}(x,z)$ for each region \cite{RM}. The flow is divided in small cells $l_x\times l_y\times l_z=2\times 1\times 2$ ($y<0$ or $y>0$), which are large enough to contain a coherent structure. The square norm of the departure to the laminar baseflow $\textbf{v}^2$ is averaged in each cell, and a criterion $\gamma$ is applied. If the average is larger than $\gamma$, the cell is considered to be turbulent, otherwise is it considered to be laminar. One can then educe each region: if two laminar cells are on top of each other, the zone $(x,z)$ is considered to be laminar. If two turbulent cells are on top of each other, the zone $(x,z)$ is considered to be turbulent. And if a laminar cell is on top of a turbulent one (or \emph{vice versa}), the zone $(x,z)$ is considered to be intermediate, or overhanging \cite{cole66}. One can further distinguish the two intermediate zones, one where a laminar cell is on top of turbulent cell and the one where a turbulent cell is on top of a laminar cell. By an abuse of language, the first intermediate zone is call the ``rear'' zone, while the second is called the ``front'' zone, since by starting the laminar zone and increasing $x$ toward the band, one enters the rear of the band, go through the turbulent zone and exists toward another laminar zone through the front of the band (similarly to what is done in \cite{ispspot}). The spatial masks are then defined by:
$I^{\rm i}(x,z)=1$ (resp. if $I^{\rm t}=1$) if $(x,z)$ belongs to any of the intermediate (resp. turbulent) zones and
$I^{\rm i}(x,z)=0$ (resp if $I^{\rm t}=0$) otherwise.

\begin{figure*}
\centerline{\includegraphics[width=17cm,clip]{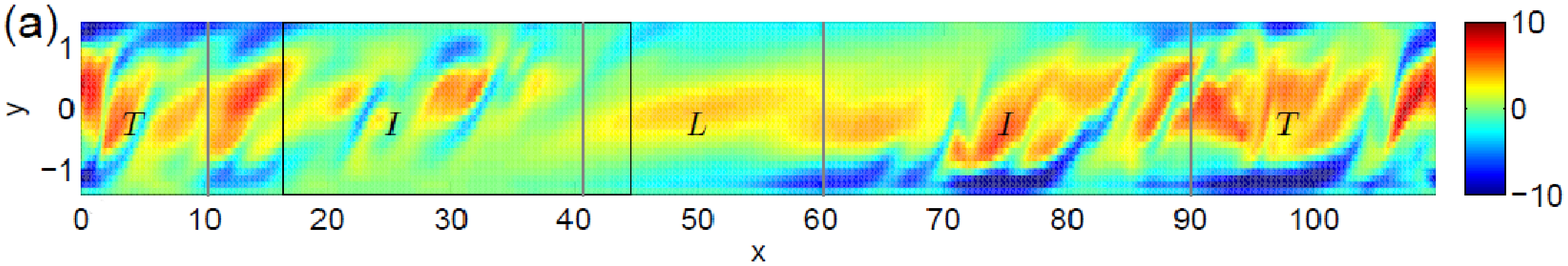}}
\centerline{\includegraphics[height=6cm,clip]{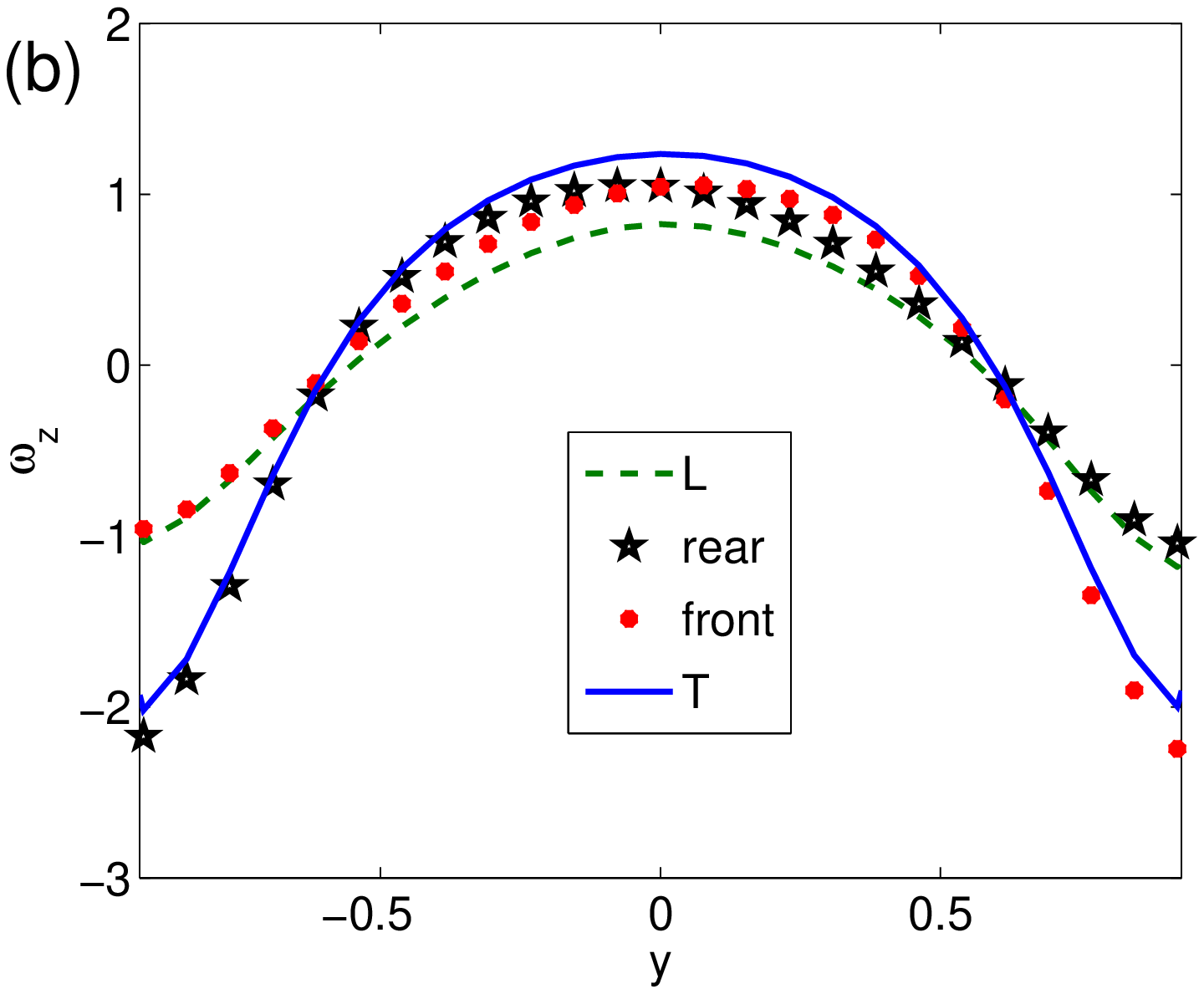}}
\centerline{{\large
\textbf{(c)}\includegraphics[width=6cm,clip]{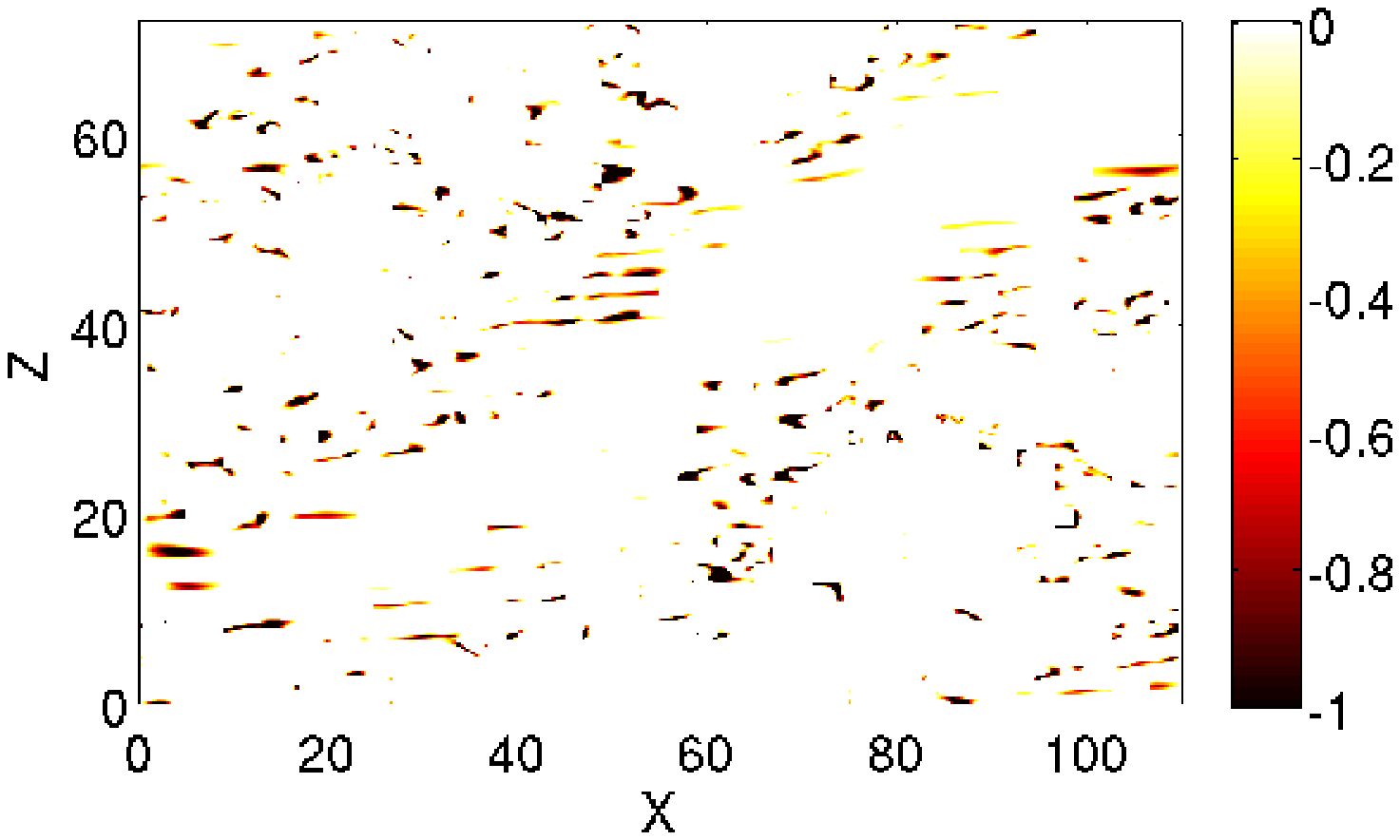}\textbf{(d)}}\includegraphics[width=6cm,clip]{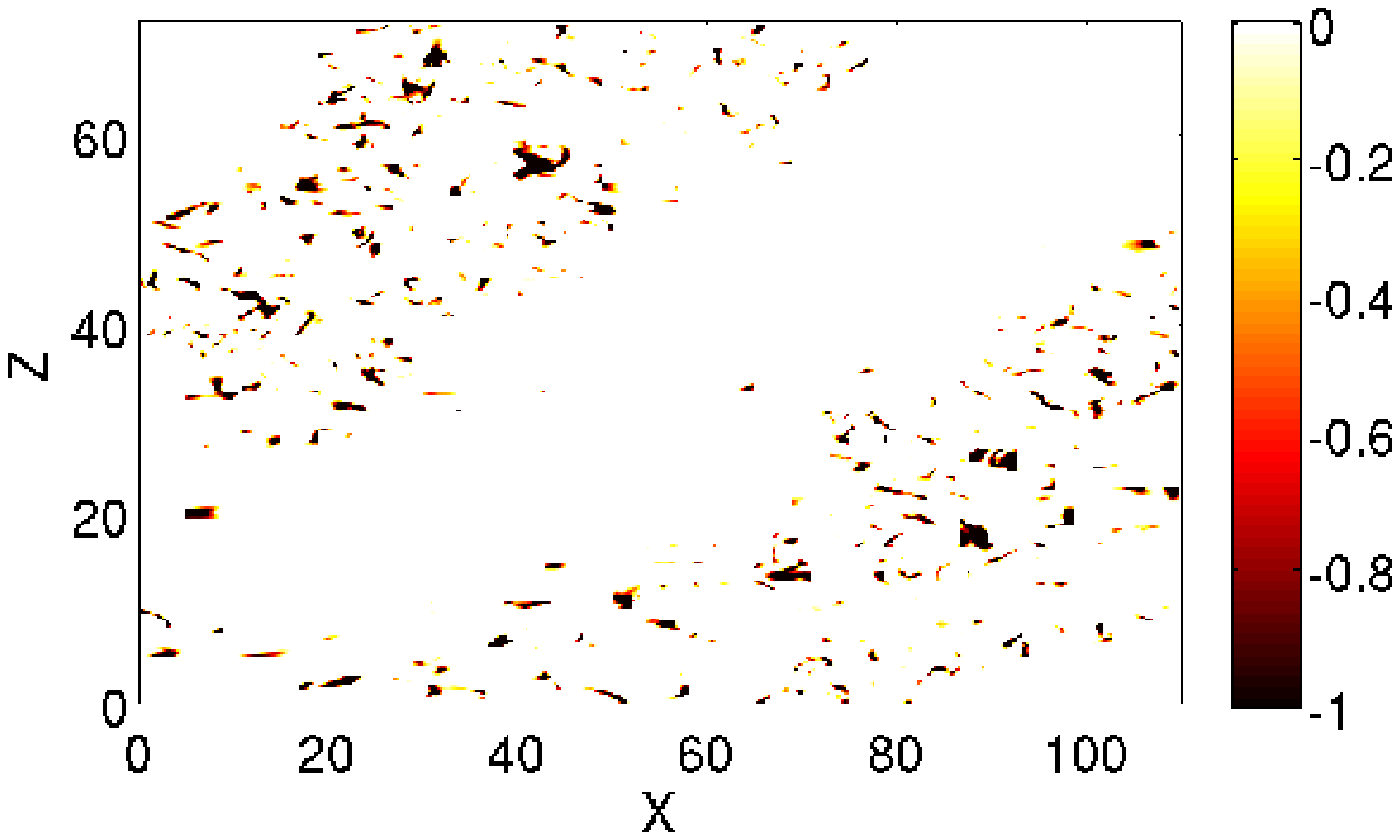}}
\caption{(a) Colour levels of the spanwise vorticity field $\omega_z$ in the $z=constant$ plane
of figure~\ref{f2}, $T=20$. The Intermediate, Turbulent and Laminar zones are indicated respectively by letters I, T and L and separated by gray lines. (b) Profiles of spanwise vorticity
conditionally averaged  in turbulent
and intermediate area. Thresholded and spatially filtered spanwise
vorticity in the $y=0$ plane: (c) $\bar{\omega}_z^{\rm i}$, (d)
$\bar{\omega}_z^{\rm t}$. (colour levels are saturated at
$-1$)}\label{f3}
\end{figure*}

By doing a conditional average of $\omega_z$ in each of these four zones (laminar, intermediate rear, turbulent, intermediate front), one obtains the profiles of figure~\ref{f3} (b). The shear $-\partial_y v_x$ dominates in $\omega_z$. The vorticity profile $\omega_z$ in the turbulent zone is positive and maximum at $y=0$, while it is negative near the walls. It has the same shape, with smaller amplitude, in the laminar zone. In the intermediate zones, the profile matches that of the turbulent zone in a half gap and that of the laminar zone in the other one. Note that since our procedure averages over the spanwise modulation of the velocity streaks, theses profiles are approximately halved compared to what can typically be found in the heart of velocity streaks (Fig.~\ref{f3} (a)). This gives a quantitative base to the observation of the properties of the vorticity field in a frozen velocity streak.

\subsubsection{The marker}

We eventually show the effect of the rolls on the vorticity field using kinematic argument on the velocity and vorticity fields. Indeed, one can approximately write, at constant $z$ inside a streak, in presence of a roll centered at $y=0$:
\begin{equation}v_y\simeq  v_y^{\rm s}(X,y)+\alpha(t)(1-y^2)^2\sin(k x)\,,\end{equation}
This description accounts for the spatial dependence of $v_y$ in a streak that contains rolls: the flow alternatively goes up and down with a wavelength $\lambda=2\pi/k$. We denote $v_x^{\rm s}$, $v_y^{\rm s}$, $\omega_z^{\rm s}$, the velocity and vorticity fields of the streaks and $\alpha(t)$ the amplitude of the perturbation. The variable $X$ corresponds to the slow
spatial dependence, on a scale of order $L_x=110$, as opposed to the fast spatial dependance of the rolls ($k\simeq 1$). The shape and amplitude of the slowly varying velocity fields in each of the zones have been computed by Barkley \& Tuckerman \cite{BT07}.

Using the incompressibility $\partial_x v_x+\partial_y v_y=0\Rightarrow v_x=\int {\rm d}x \partial_y v_y$ of the two dimensional roll, this
dependence leads to:
\begin{equation}v_x\simeq v_x^{\rm s}(X,y)+\frac{4\alpha(t)}{k}
y(y^2-1)\cos(kx)\,,\end{equation}
for $v_x$ inside a streak containing a roll. This spatial dependence describes the field $v_x$ computed in our DNS very well (Fig.~\ref{f2} (a,d)).

 The wall normal dependence of the perturbation in $v_y$ is that of
the first function of the orthogonal basis fitting the boundary
condition for $v_y$. The wall normal dependence of the perturbation in $v_x$ is that of the
second function of the basis fitting the boundary conditions for
$v_x$. These bases date from early studies of thermal convection \cite{conv}. This type of polynomial descriptions are commonly used to describe the velocity field \cite{Jimenez91}. Even at lowest order, these bases can
approximate most of the wall normal dependence of the flow \cite{m}.
 These two bases of function will be used and detailed more
extensively in the second part of the article \cite{isp2}.

One can then approximate the spanwise vorticity by:
\begin{equation}
\omega_z\simeq  \omega_z^{\rm s}(X,y)+\alpha(t) \cos(kx)\left(k(1-y^2)^2
+\frac{4-12y^2}{k} \right)\,,\label{eqvrt}\end{equation}
 with $\omega_z^{\rm s}\simeq -\partial_y v_x^{\rm s}$. The $y$ dependance of the slowly varying vorticity field can be seen in figure~\ref{f3} (b) (with a factor $1/2$). The additive
perturbation of $\omega_z$ has the same shape as the spanwise vorticity, with a fast modulation in $x$. Both the
contributions of $\partial_y v_x$ and $\partial_x v_y$ have
the same sign. In the midplane, the
additive perturbation leads to a change of sign of the spanwise
vorticity everywhere $\cos(x)$ is negative.

  One can see that detecting $\omega_z<0$ near $y=0$ is equivalent to
detecting the rolls. Negative spanwise vorticity around $y=0$
can be used as a marker of the developing instability. The field
$\omega_z^{\rm th}$, spanwise vorticity thresholded at zero:
\begin{equation}
\omega_z\le 0 \Leftrightarrow \omega_z^{\rm th}=\omega_z\, , \quad \omega_z> 0 \Leftrightarrow \omega_z^{\rm th}=0\,
\end{equation}
is used. In the range $-0.5\lesssim y\lesssim 0.5$, the field $\omega_z^{\rm th}$ is non zero if the rolls appear and  zero
if it is not present.

  One might think that such spanwise vorticity results from the tilting of wall normal vorticity created by the streak instability \cite{HKW,W} or from tilting of the streamwise vortices. This is not the case, such an effect has never been reported. Streamwise vorticity and thresholded spanwise vorticity are decorrelated. Indeed using velocity fields computed in this study, one finds (see \cite{rmq} for the definition of brackets) :
\begin{equation}\left|
\frac{\langle (\omega_x-\langle \omega_x\rangle_{x,z})(\omega_z^{\rm th}-\langle \omega_z^{\rm th}\rangle_{x,z}) \rangle_{x,z}}{\sqrt{\langle (\omega_x-\langle \omega_x\rangle_{x,z})^2\rangle_{x,z}\langle (\omega_z^{\rm th}-\langle \omega_z^{\rm th}\rangle_{x,z})^2\rangle_{x,z}}}\right|\lesssim 0.05\,.
\end{equation}
The computation is performed in the plane $y=0$.
  This can be easily understood. The examination of the vorticity evolution equation shows that the tilting of $\omega_y$ into $\omega_z$ would root from the wall normal shear of spanwise velocity, through the term $\omega_y \partial_y v_z$. The first non zero term in the centerline results from the large scale flow around the bands, leading to $\partial_y v_z\propto O(10^{-2})$ \cite{BT07}. This is much smaller than the wall normal shear of streamwise velocity (of order $1$) responsible for the creation of streamwise vortices. The tilting of streamwise vortices is even less likely. Indeed, it would root from $\omega_x \partial_x v_z$. Again, the first non zero contribution results from the large scale flow around the bands. One then has $\partial_x v_z\propto O(10^{-4})$, since the streamwise scale of variation of the large scale flow is the wavelength of the bands.

 The laminar/intermediate/turbulent discrimination procedure is used to mask the thresholded spanwise
vorticity. It yields two fields $\omega_z^{\rm th, i}=I^{\rm
i}\omega_z^{\rm th}$ (figure~\ref{f3} (c)) and $\omega_z^{\rm
th, t}=I^{\rm t}\omega_z^{\rm th}$ (figure~\ref{f3} (d)). They allow one
to monitor the rolls in these zones only. Note that the vorticity appears uniformly in the band: this differs from the case of the early spots, which contain two core of production of spanwise vorticity \cite{ispspot}.

\section{Measurements \label{mes}}

  The thresholded (and masked) vorticity fields are now used for the systematic measurement of the characteristics of the rolls: their size and as their advection velocity. We follow the approaches proposed in the study of the spots \cite{ispspot}.

\subsection{Lengthscale measurements\label{span}}

\begin{figure}
\centerline{\includegraphics[width=5.5cm,clip]{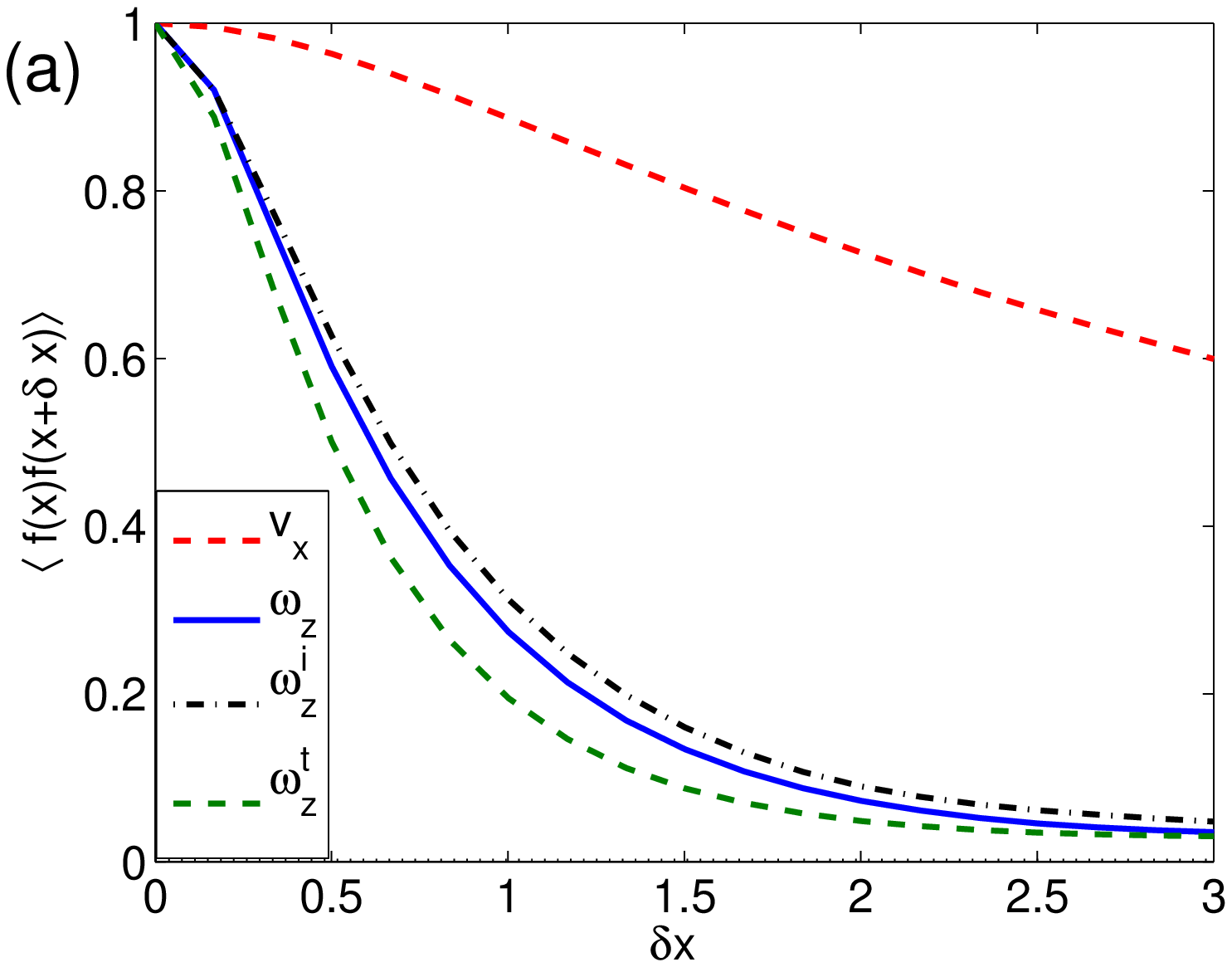}\includegraphics[width=5.5cm,clip]{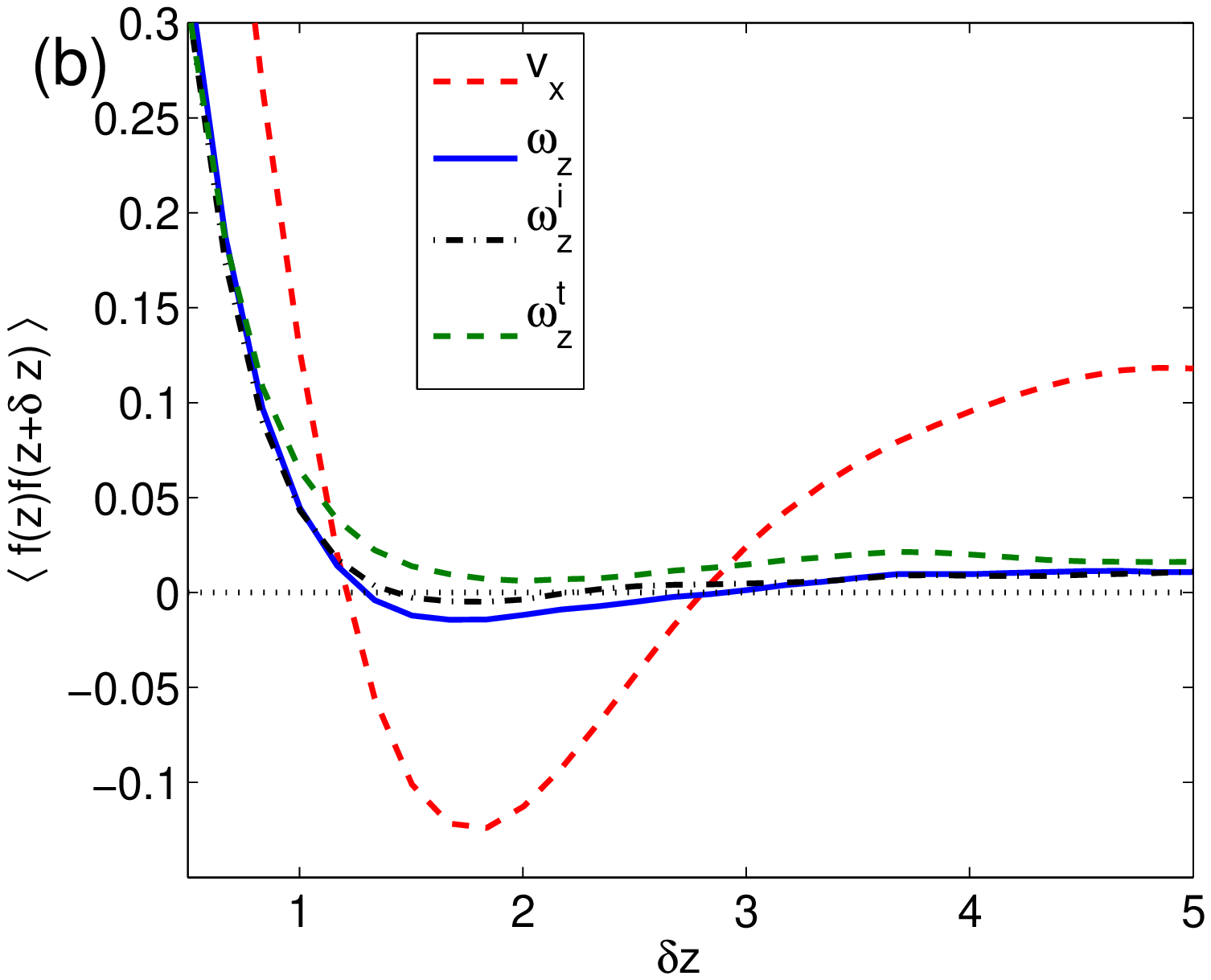}}
\caption{Correlation function in the $x$ (a) and $z$ (b) direction of streamwise velocity $v_x$, thresholded vorticity field $\bar{\omega_z}$
and masked $\bar{\omega}_z^{\rm i}$ and $\bar{\omega}_z^{\rm t}$, averaged in time.}
\label{lgsc}
\end{figure}

  In order to determine the characteristic sizes of the rolls in both streamwise and spanwise direction, and compare them to that of the velocity streaks, we turn to normalised correlation functions : $\langle f(x,z)f(x+\delta x,z+\delta z)\rangle$ \cite{rmq}. The field $f$ can be the masked thresholded vorticity $\omega_z^{\rm th}$, $\omega_z^{{\rm th}, {\rm i},{\rm t}}$, or the  streamwise velocity $v_x$. The correlation functions are computed at $y=0$ where the perturbation to the velocity streaks is maximum. They
are time averaged, a minimum of ten samples is necessary for
convergence. The envelope of the correlation functions gives to the
characteristic correlation length of the field. The modulation
indicates periodicity.

  The streamwise correlation functions of the thresholded
vorticity fields decrease exponentially (Fig.~\ref{lgsc}
(a)). The characteristic size is estimated from the value $\delta x$
for which the tangent at $\delta x=0$ crosses the $\langle f(x)
d(f+\delta x)\rangle =0$ line. This characteristic size is also the inverse of the slope at $\delta x=0$. This yields a coherence length ranging from $1$
($\omega_z^{\rm th, t}$) to $1.5$ ($\omega_z^{\rm th, i}$). This puts figures on the difference of lengthscales seen in the visualisations \cite{ispspot}.
Since the spanwise vorticity is negative on less than half a wavelength, this gives wavelength of more than $2$ and $3$, consistent with the constatation of rolls of size $5$ (Fig.~\ref{f2}). The ratio of coherent length can be trusted as
the ratio of typical length scales of the perturbations in the
intermediate and turbulent zone. The velocity field has a much
longer coherence length, of order $10$. It corresponds to the velocity streak
coherence length. This shows the scale separation between the
velocity streaks in the bands and the perturbations and places this secondary instability in the same framework as the study of the growth of the spots \cite{PRL}.

  The spanwise dependence of $\omega_z^{{\rm th}, {\rm i},{\rm t}}$
can be examined in the time averaged correlation function as well: the correlation function as a function of
$\delta z$, at $\delta x=0$, is displayed in figure~\ref{lgsc} (b). The figure is zoomed
around $\langle f(z)f(z+\delta z)\rangle=0$. The slope at the origin and the first zero of each
correlation function is an estimate of the thickness of velocity
streaks (for $v_x$) and vorticity (for $\omega_z^{\rm th}$). One finds
a spanwise coherence length for $ \omega_z^{{\rm th}, {\rm i},{\rm t}}$
of $1$, slightly smaller than that of $v_x$. In both the intermediate and turbulent zone, spanwise vorticity $\omega_z^{\rm th}$ is
concentrated in locations slightly smaller than the turbulent streaks. The
modulation of the correlation function of $v_x$ shows the
spanwise modulation of the velocity streaks. The vorticity fields $\omega_z^{\rm th}$ and
$\omega_z^{\rm th, t}$ bear no trace of long range correlation
and periodicity, whereas $\omega_z^{\rm th, i}$ has a small trace
of periodicity. This is consistent with the constatation that perturbations
develop independently in each velocity streak (Fig.~\ref{f3} (c,d)). This background is quasi periodical
and can leave a hint of periodicity in the perturbation field
$\omega_z^{\rm th}$.

\subsection{Advection of perturbations\label{advpert}}

  In this section, we go beyond the visualisation of figure~\ref{f2} and measure the advection velocity of the rolls.
This advection can be seen in a video of the thresholded spanwise vorticity $\omega_z^{\rm th,i}$ zoomed in an intermediate zone, available in the supplementary material. The
$y=0$ plane is chosen. The spanwise vorticity remains
coherent in time. It clearly moves in the increasing $z$ direction. However, the
advection in the streamwise direction is not as clearly identifiable
as in the spanwise direction. In this section, we investigate this matter more
thoroughly using the systematic advection velocity measurement
procedure proposed in the study of the spots \cite{ispspot}.

\subsubsection{The processing procedure\label{gen}}

  The advection is characterised by its velocity. A direct
measure of velocity $\mathbf{c}=c_x\mathbf{e}_x+c_z\mathbf{e}_z$ for
each perturbation is
heavy and cumbersome in plane Couette flow. The measurement is therefore
automated \emph{via} an image correlation approach. This is similar to Particle Image Velocimetry
algorithms, measurement techniques proposed over thirty years ago which have become commonplace in the last twenty years \cite{AnnRevPIV}. The non masked thresholded spanwise vorticity $\omega_z^{\rm th}$ is used for measurement.

The measurement procedure is as follow. A given $x-z$ plane is divided in squares of
size $2\times 2$. This size is small enough to capture the small
scale details of the flow, and large enough to contain enough
information \cite{RM}. Then for each cell, at time $t$, the
correlation to a cell shifted by $\Delta z$ at time $t+\delta t$ is
computed:
\begin{align}\notag
C_{x_0,z_0,y,t}(\Delta z,\delta t)=\\\int_{x_0,z_0}^{x_0+2,z_0+2}{\rm d}x{\rm d}z\, \tilde{\omega}(x,y,z,t)\tilde{\omega}(x,y,z+\Delta z,t+\delta t)\,,
\end{align}
where $x_0,z_0$ denominates the position of the ``lower left'' corner of the cell
. One has $x_0=2m_x$, $z_0=2m_z$ with $m_{x,z}$ integers. $\tilde{\omega}$ denominates $\left(\omega_z^{\rm th}-\langle
\omega_z^{\rm th} \rangle\right)/(\langle(\omega_z^{\rm th}-\langle
\omega_z^{\rm th} \rangle)^2 \rangle)^\frac12$. Here $\langle .\rangle$ is the average over the relevant $2$ by $2$ square.

  For each cell at $x_0,z_0$ at each time $t$,
for a given $\delta t$, the shift $\Delta z$ maximising
$C_{x_0,z_0,y,t}(\Delta z)$ is computed. The corresponding velocity
for the cell is $c_z(y)=\Delta z/\delta t$. A field of advection velocity
is obtained. An example of $c_z$ at $y=0$ is displayed in colour levels in figure.~\ref{advspd_}, (a). The partial
coarse graining in the $(x,z)$ plane is visible.

  The field of advection velocity is smoothed before the analysis. Indeed, when little to no marker are present in the
cell, the correlation function has no meaning, maxima can arise
for any value. The laminar zone is the most sensitive one, due to
the very small density of markers (spanwise vorticity). Given the translational invariance along the band, an average in the $z'$ direction is performed (figure~\ref{advspd_} (b)). The
computation depends weakly on $\delta t$ provided it remains in the $1\le
\delta t\le 4$ range. For higher values of $\delta t$ coherence is lost. The trackers can be advected to far away in
$x$ and $z$. The tracker (vorticity) evolves in time and is not
steady enough in shape.

  This procedure can be applied in a similar fashion
for increments of space  in the streamwise direction $\Delta x$.
The two cases are examined separately. The streamwise direction requires a little more processing. This is why the two computations are decoupled.

\subsubsection{spanwise advection}

\begin{figure}
\centerline{\includegraphics[width=7cm,clip]{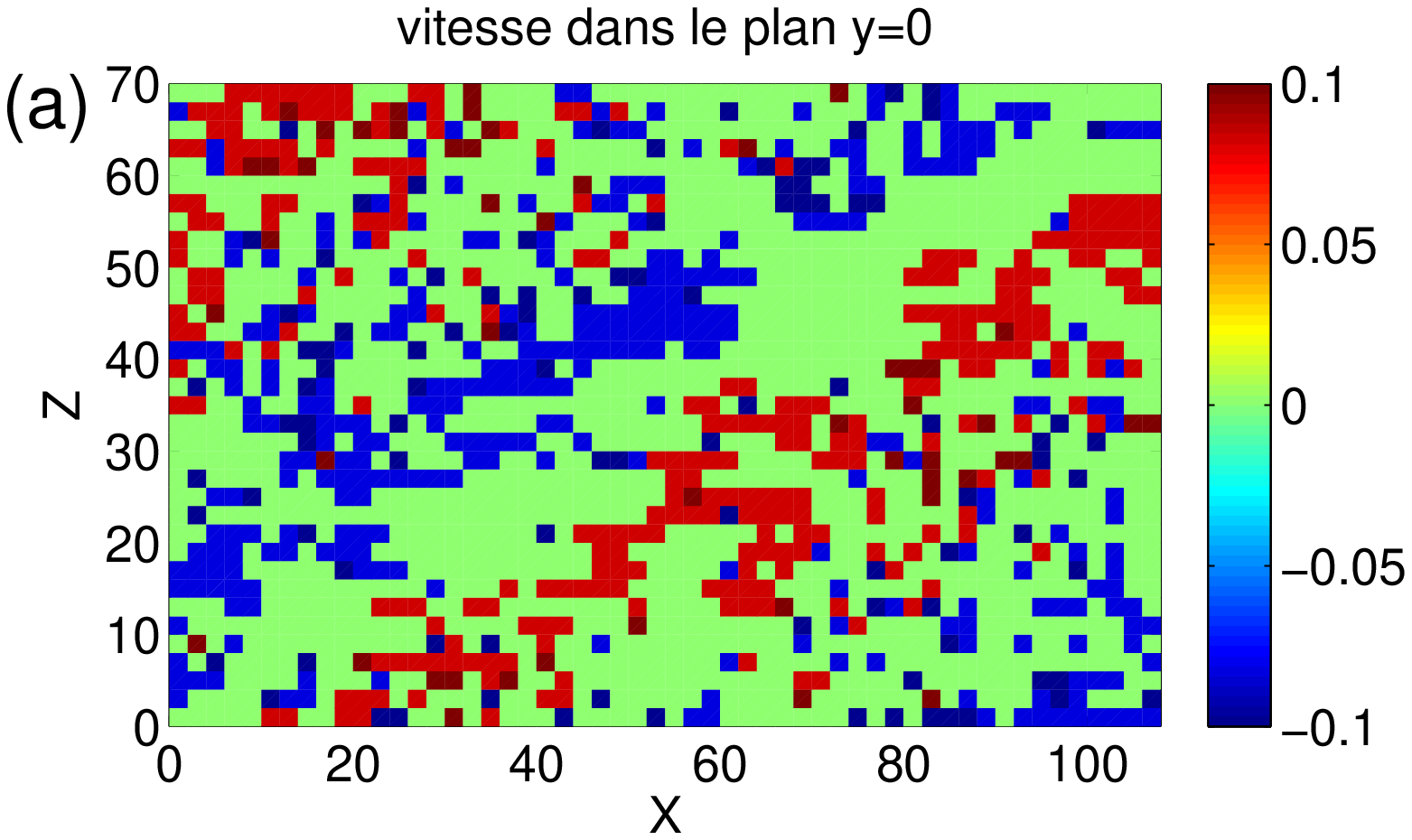}\includegraphics[width=7cm,clip]{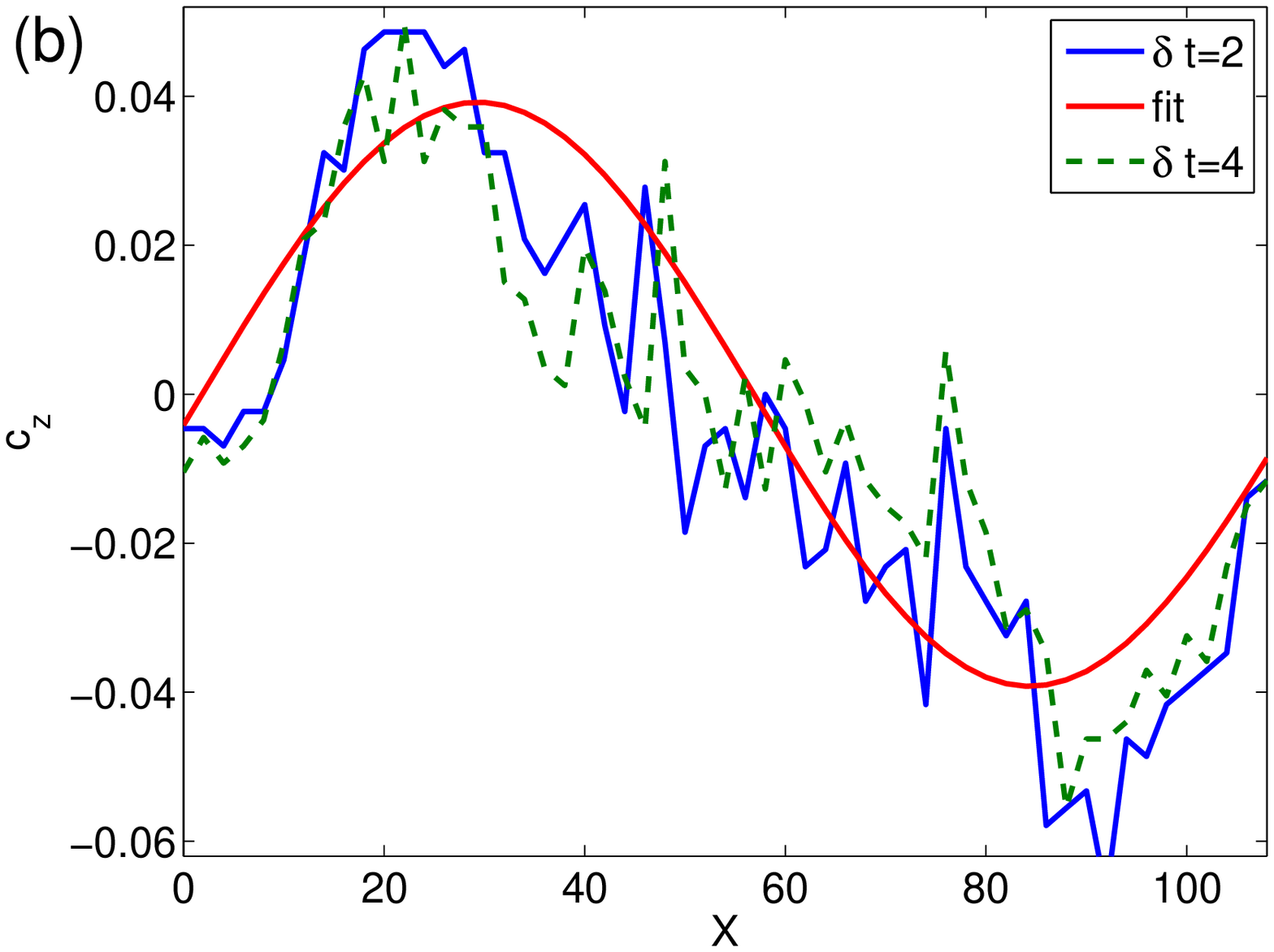}}
\centerline{\includegraphics[width=7cm,clip]{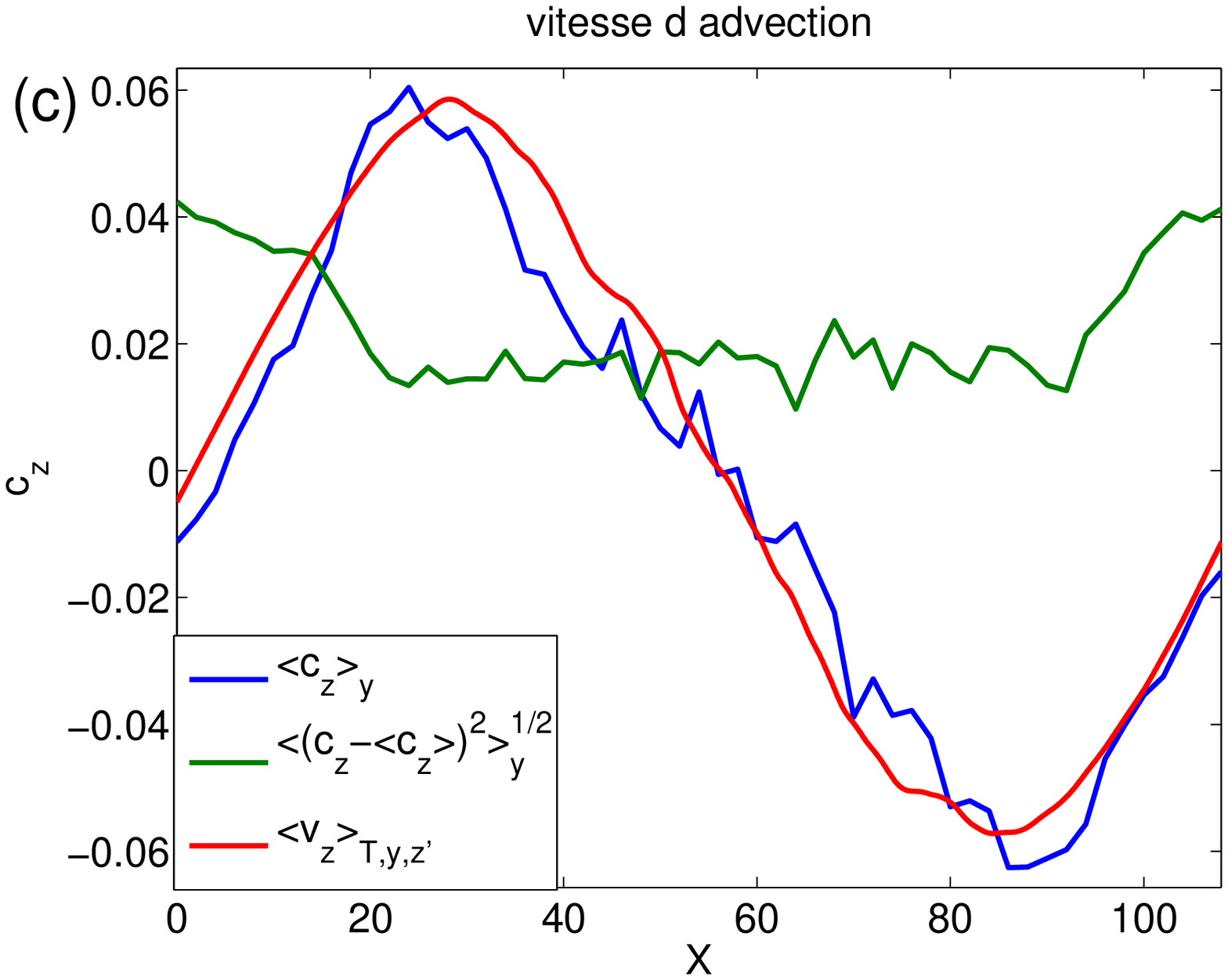}
\includegraphics[width=7cm,clip]{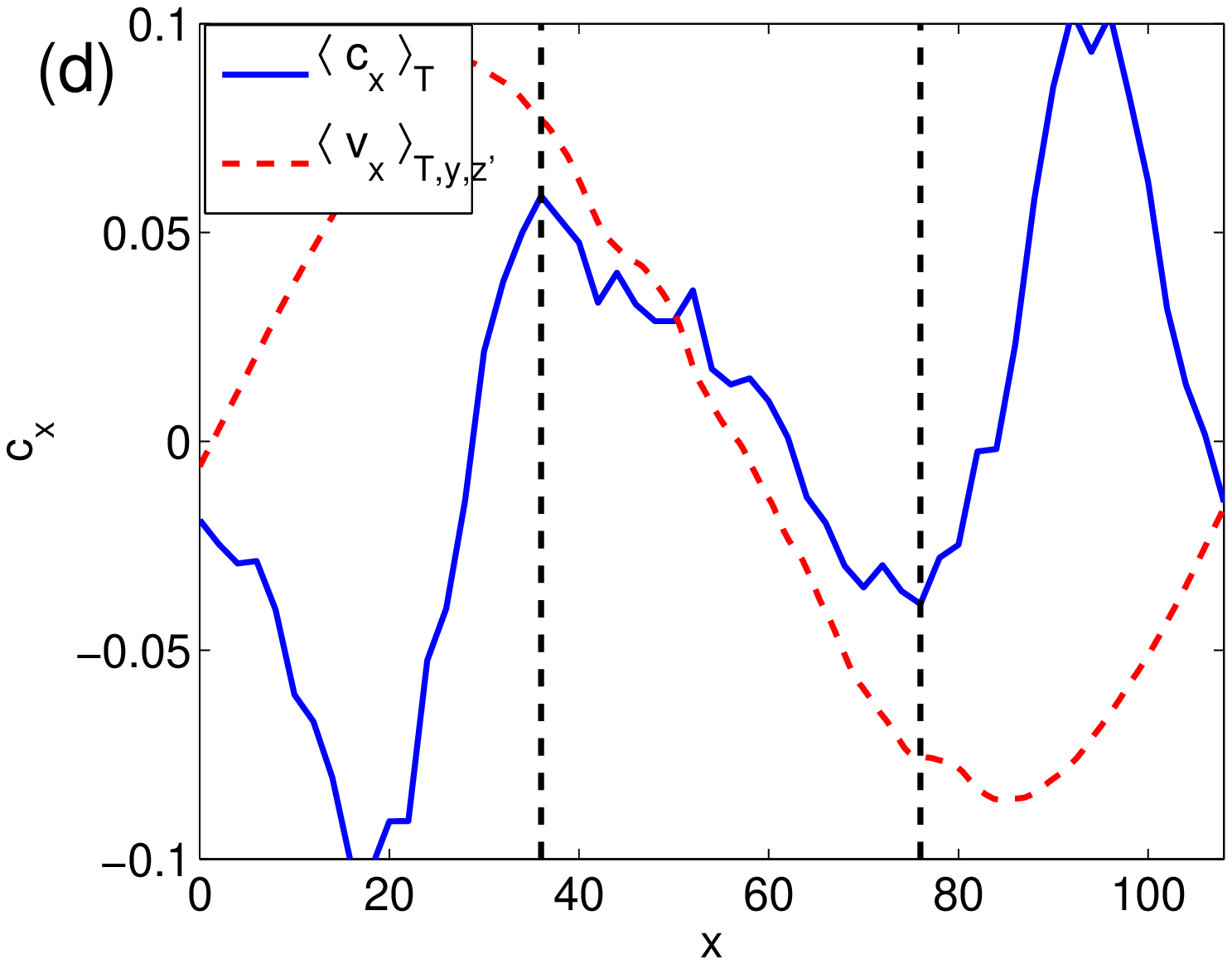}}
\caption{(a) : Field of spanwise advection velocity $c_z$, in the case of band, at $y=0$.
(b): Advection velocity $c_z$ averaged over the diagonal at a given time at
$y=0$ for two time steps $\delta t=2$ and $\delta t=4$ and a sinusoidal fit.
(c): Average and fluctuation of the advection velocity $c_z$ over the diagonal
and the wall normal direction at a given time, compared to the spanwise large scale flow.
(d) Comparison of the measured streamwise advection velocity, $\delta t=1$, at $y=0$
and the averaged large scale flow in the band.}
\label{advspd_}
\end{figure}

  The spanwise advection velocity $c_z$ as a function of $x$ is displayed in figure~\ref{advspd_} (b,c). The instantaneous result is coherent (Fig.~\ref{advspd_} (b)). The wall normal and time ($200 h/U$) average of $c_z$ is nearly sinusoidal (Fig.~\ref{advspd_} (c)) and is perfectly matched by the spanwise component of the large scale flow (Fig.~\ref{advspd_} (c)). As mentioned in the introduction, the large scale flow is obtained by the $y$, $z'$ and time average of the spanwise velocity field. This shows that, in the spanwise direction, the rolls travel at the velocity of the wall-normal averaged large scale flow. The standard deviation over $y$ of the advection velocity is computed as well. It shows that $c_z$ depends very weakly on $y$, except in the laminar region, where there are very few markers.

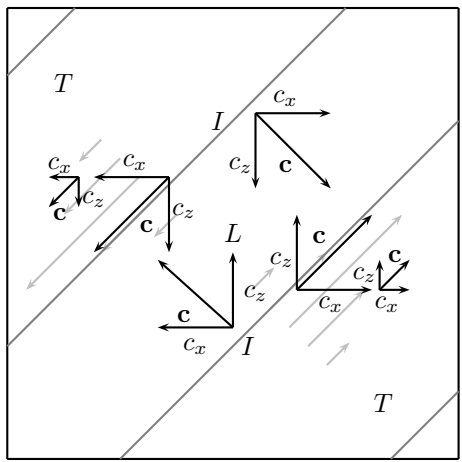
\begin{figure}
\centerline{
\begin{pspicture}(7,7)
\psline[linecolor=lightgray]{->}(1.75,4.75)(1.45,4.45)
\psline[linecolor=lightgray]{->}(2,4.5)(1.25,3.75)
\psline[linecolor=lightgray]{->}(2.25,4.25)(0.75,2.75)
\psline[linecolor=lightgray]{->}(2.5,4)(1.75,3.25)
\psline[linecolor=lightgray]{->}(2.75,3.75)(2.45,3.45)
\psline[linecolor=lightgray]{->}(4.25,2.25)(5.75,3.75)
\psline[linecolor=lightgray]{->}(4.5,2)(5.25,2.75)
\psline[linecolor=lightgray]{->}(4.75,1.75)(5.05,2.05)
\psline[linecolor=lightgray]{->}(4,2.5)(4.75,3.25)
\psline[linecolor=lightgray]{->}(3.75,2.75)(4.05,3.05)
%fin
\psline{}(0.5,0.5)(0.5,6.5) \psline{}(0.5,6.5)(6.5,6.5)
\psline{}(6.5,6.5)(6.5,0.5) \psline{}(0.5,0.5)(6.5,0.5)
\psline[linecolor=gray]{}(0.5,2)(5,6.5) \psline[linecolor=gray]{}(2,0.5)(6.5,5) \rput(3.5,3.5){$L$}
\rput(3.3,5){$I$} \rput(3.7,2){$I$} \rput(1.25,5.5){$T$}
\rput(5.5,1.25){$T$}
%fin
\rput(1.65,3.95){$c_z$} \rput(1.2,4.4){$c_x$}
\rput(1.2,3.75){$\mathbf{c}$} \psline{->}(1.45,4.25)(1.05,4.25)
\psline{->}(1.45,4.25)(1.45,3.85) \psline{->}(1.45,4.25)(1.05,3.85)
%fin
\rput(4.2,5.3){$c_x$} \rput(3.6,4.4){$c_z$}
\rput(4.2,4.4){$\mathbf{c}$} \psline{->}(3.8,5.1)(4.8,5.1)
\psline{->}(3.8,5.1)(3.8,4.1) \psline{->}(3.8,5.1)(4.8,4.1)
%fin
\rput(2.2,4.45){$c_x$} \rput(2.85,3.8){$c_z$}
\rput(2.35,3.6){$\mathbf{c}$} \psline{->}(2.65,4.25)(1.65,4.25)
\psline{->}(2.65,4.25)(2.65,3.25) \psline{->}(2.65,4.25)(1.65,3.25)
%fin
\psline{->}(4.35,2.75)(5.35,2.75) \psline{->}(4.35,2.75)(4.35,3.75)
\psline{->}(4.35,2.75)(5.35,3.75) \rput(4.15,3.15){$c_z$}
\rput(4.8,2.55){$c_x$} \rput(4.65,3.45){$\mathbf{c}$}
%fin
\psline{->}(5.45,2.75)(5.85,2.75) \psline{->}(5.45,2.75)(5.45,3.15)
\psline{->}(5.45,2.75)(5.85,3.15) \rput(5.55,2.55){$c_x$}
\rput(5.25,2.95){$c_z$} \rput(5.65,3.2){$\mathbf{c}$}
%fin
\psline{->}(3.5,2.25)(2.5,2.25) \psline{->}(3.5,2.25)(3.5,3.25)
\psline{->}(3.5,2.25)(2.5,3.15) \rput(3,2){$c_x$}
\rput(3.8,2.7){$c_z$} \rput(2.85,2.4){$\mathbf{c}$}
%fin
\psline[linecolor=gray]{}(0.5,5.6)(1.4,6.5)
\psline[linecolor=gray]{}(5.6,0.5)(6.5,1.4)
\end{pspicture}
}
\caption{Sketch of the flow (based on sketch~\ref{f1_} (b)), indicating the band, the Laminar, Intermediate and turbulent zones, as well as both components $(c_x,c_z)$ and the resulting advection velocity $\mathbf{c}$ in each zone.}
\label{sketchres}
\end{figure}

  The spatial dependence of $c_z$ is summed up in the sketch of the $x-z$ plane (Fig.~\ref{sketchres}) based on sketch~\ref{2d2c} (b). The turbulent band is indicated, as well as the Turbulent, Intermediate and Laminar zones. The direction of the spanwise advection velocity is indicated by the $c_z$ vector. Since $c_z=0$ in the middle of the turbulent and pseudo-laminar zone, no vector is drawn. $c_z$ is positive in one intermediate zone ($10\lesssim x\lesssim 40$) and negative in the other one ($60\lesssim x \lesssim 100$).

\subsubsection{Streamwise advection}

  We then apply the same procedure to measure $c_x$ (Fig.~\ref{advspd_} (d)). Due to stronger fluctuations than in the measurement of $c_z$, time
average is absolutely necessary. Good agreement is found between $c_x$ and
the streamwise component of the large scale flow around the turbulent area ($35 \lesssim x \lesssim 75$). However,
 a different picture is found for $0\lesssim x \lesssim
35,\,75\lesssim x\lesssim 110$, in the laminar region and part of the intermediate regions. The advection velocity strongly differs from the large scale flow, and even changes sign. In
that case the vorticity is no longer advected along the band (taking
into account $c_z$) but toward the laminar zone. This may very well be an effect of the laminar baseflow on perturbation
localised in the intermediate zones. Indeed, perturbations in the front intermediate
zone ($y>0$ part of the flow) see a positive effective velocity,
while perturbations in the rear intermediate zone ($y<0$ part of the flow)
see a negative effective velocity. The contribution of the laminar baseflow balances out
inside near the turbulent zone.

  The direction of the streamwise component of the advection velocity is included in the sketch~\ref{sketchres}. It is indicated by the $c_x$ vectors. One has $c_x=0$ in the middle of the turbulent and laminar regions, hence the absence of vector. In the intermediate region, the two directions of advection the are indicated by vectors of opposite signs, one near the turbulent region and one near the laminar region. The vectorial advection velocity $\mathbf{c}=c_x \mathbf{e}_x+c_z\mathbf{e}_z$ is added. One can see that around the turbulent region, the spanwise vorticity is advected along the turbulent band, whereas around the laminar region, the perturbations are advected toward the laminar region, in a direction nearly orthogonal to the band.

\section{Discussion \label{concl}}

  In this article, we identified the formation of rolls in the shear layers of velocity streaks of the laminar-turbulent oblique bands of plane Couette flow, which lead to spanwise vorticity. These rolls are very similar to those found in Hagen--Poiseuille flow \cite{DWK,SK}, plane Poiseuille flow \cite{ATK} or in spots of PCF \cite{ispspot}. We justified quantitatively the use of a criterion based on the sign of the vorticity in the midplane to systematically detect these events and used it to perform measurements. Correlation functions have been used to measure the lengthscale of said rolls. This stressed on the scale separation between the large scale flow and the vorticity as well as the localisation of the vorticity inside the streaks. We extended a method to measure the advection velocity of the vorticity used in the study of the spots \cite{ispspot} to the case of the bands. It showed that the advection velocity of the rolls quantitatively matched the large scale flow along the bands, except in the laminar region where vorticity is advected away from the bands. The onset of the secondary instability creating the rolls, as well as its convective or absolute nature will be investigated in the second part of the article (a preliminary study can be found in \cite{ETC}).

  We will focus on equivalent of the sustainment cycles of the puffs of pipe flow (see \cite{DWK,SK}) in the second part of the article. It will link the advection along the bands and a possible feed back mechanism. The feed back found in the sharp trailing edges of puffs rooting in an inflectional instability is likely in the same class of mechanism \cite{HDAS}. We focus here on the possible effect of the advection of vorticity away from the band on the distance between two bands. DNS of PCF showed that, in time average, there was a precise force balance between dissipation and the advection by the laminar baseflow. This gave a relation between the Reynolds number and the wavelengths of the bands and this mechanism may be at the source of the distance between two bands \cite{BT07}. Similarly, The effect of leading edge on the trailing edge of the next puff was proposed to explain the distribution of distance between two puffs centered around a well defined average distance \cite{sam}. If the puffs are too close, the inflexion of the streamwise velocity profile at the leading edge is drastically reduced \cite{HDAS}. The advection of the spanwise vorticity of the rolls back toward the laminar zone where it is dissipated, instead of feeding turbulence, may be the instantaneous version of such an interaction in PCF.  This advection-dissipation mechanism would also be the instantaneous version of the average force budget. In that matter, PCF differs from the case of the slugs where the advection of the azimuthal vorticity mainly fed the extension of turbulence downstream of the slug \cite{DWK}. It can be argued that this is partially the effect of the centro-symmetry of PCF, and partially the effect of the difference of Reynolds number. Indeed, dissipation is a fundamental part of the small scale dynamics at these Reynolds numbers \cite{PRL}. Without the band structure, turbulence in PCF naturally decays if $R\lesssim 415$ \cite{PM,DSL}. Meanwhile, the slug regime corresponds to $R>R_{\rm t}$, for those Reynolds number, the advection of the rolls toward laminar flow may contribute to the unlimited streamwise extension of the spots.

  The strong similarities between Pipe and Couette flows in that matter motivates further studies aimed at understanding the very complex mechanisms behind the laminar-turbulent coexistence.

\section*{acknowledgments}
The author acknowledges discussions with Y. Duguet, P. Huerre and P. Manneville.

\end{document}